\documentclass[aps,amsmath,twocolumn,showpacs]{revtex4-1}

\usepackage{epsfig}
\usepackage{graphics}
\usepackage{latexsym}
\usepackage{amsmath}
\usepackage{amssymb}
\usepackage{rotating}
\usepackage{subfigure}
\usepackage{bm}
\usepackage{color}
\usepackage{blindtext}
\usepackage{lineno}
\begin{document}
\title{Predictions for the isolated prompt photon production at the LHC at $ \sqrt s= $13 TeV}

\author{Muhammad Goharipour}
\email{m.goharipour@semnan.ac.ir}

\author{Hossein Mehraban}
\email{hmehraban@semnan.ac.ir}

\affiliation{Faculty of Physics, Semnan University, Semnan P.O. Box 35131-19111, Semnan, Iran}

\date{\today}

\begin{abstract}

The prompt photon production in hadronic collisions has a long history of providing information on the
substructure of hadrons and testing the perturbative techniques of QCD. Some 
valuable information about the parton densities in the nucleon and nuclei, especially of the gluon,
can also be achieved by analysing the measurements of the prompt photon production cross section
whether inclusively or in association with heavy quarks or jets.
In this work, we present predictions for the inclusive isolated prompt photon production in $ pp $ collisions
at center-of-mass energy of 13 TeV using various modern PDF sets. The calculations are presented
both as a function of photon transverse energy $ E_\textrm{T}^\gamma $ and 
pseudorapidity $ \eta^\gamma $ for the ATLAS kinematic coverage. We also study in detail the theoretical uncertainty
in the cross sections due to the variation of the renormalization, factorization and fragmentation scales.
Moreover, we introduce and calculate the ratios of photon momenta 
for different rapidity regions and study the impact of various input PDFs on such quantity.
\end{abstract}


\maketitle


\section{Introduction}\label{sec:one} 
From past to present, prompt photon production at hadron colliders
has undergone very impressive experimental~\cite{Albajar:1988im,Alitti:1992hn,Abbott:1999kd,Abazov:2001af,Abazov:2005wc,Acosta:2002ya,Acosta:2004bg,Aaltonen:2009ty,Luca:2016rvl,Khachatryan:2010fm,Chatrchyan:2011ue,Aad:2010sp,Aad:2011tw,Aad:2013zba,Aad:2016xcr,Adler:2006yt,Adare:2012yt} and theoretical~\cite{Aurenche:1983ws,Aurenche:1987fs,Owens:1986mp,Aurenche:1989gv,Baer:1990ra,Berger:1990et,Gordon:1994ut,Cleymans:1994nr,Aurenche:1998gv,Skoro:1999rg,Fontannaz:2001ek,Catani:2002ny,Bolzoni:2005xn,Aurenche:2006vj,Lipatov:2005wk,Baranov:2008sr,Schwartz:2016olw,Odaka:2015uqa,Lipatov:2016wgr,Jezo:2016ypn,Kohara:2015nda,Campbell:2016lzl,Goharipour:2017uic} developments.
The experimental measurements covers a large domain of center-of-mass energy
and also a wide range of photon transverse energy $ E_\textrm{T}^\gamma $. The prompt photon production cross section at the LHC~\cite{Khachatryan:2010fm,Chatrchyan:2011ue,Aad:2010sp,Aad:2011tw,Aad:2013zba,Aad:2016xcr} 
has a significantly higher magnitude when compared to the Tevatron~\cite{Abbott:1999kd,Abazov:2001af,Abazov:2005wc,Acosta:2002ya,Acosta:2004bg,Aaltonen:2009ty,Luca:2016rvl}. It is also much larger than 
the photoproduction cross section at HERA~\cite{Aaron:2007aa,Chekanov:2009dq,Abramowicz:2013vfa}. 
By definition, ``prompt photons" are those photons that come from the collision of two primary
partons in the protons, i.e. photons not originating from hadron decays. The study of such
photons provides a probe of perturbative Quantum Chromodynamics (pQCD)
and measurement of their production cross sections, because of the sensitivity of 
the process to the gluon content of the nucleon, can provide useful information about the
gluon parton distribution function (PDF)~\cite{Aurenche:1988vi,Vogelsang:1995bg,Ichou:2010wc,d'Enterria:2012yj,Aleedaneshvar:2016dzb}.
The associated production of prompt photons and heavy quarks, where the heavy
quarks are either charm or bottom, can also provide a powerful tool for
searching the intrinsic heavy quark components of the nucleon~\cite{Brodsky:2015fna,Rostami:2016wqa,Rostami:2016dqi}.
Moreover, a better understanding of prompt photon production is essential
to have accurate QCD predictions for physical processes for which 
the prompt photons represent an important background such as diphoton 
decays of the Higgs boson~\cite{Ball:2007zza,Aad:2014eha,Heng:2012at,Jia-Wei:2013eea}.

Inclusive prompt photon production consists of two types of photons: direct
and fragmentation photons~\cite{Catani:2002ny}. Direct photons are those produced predominantly 
from initial hard scattering processes of the colliding quarks or gluons.
Fragmentation photons are produced as bremsstrahlung emitted by a
scattered parton, from the fragmentation of quarks and gluons.
In this way, the fragmentation contribution of the inclusive prompt photon production
is expressed as a convolution of the hard parton spectra with the nonperturbative fragmentation functions (FFs).
An isolation requirement is used to reject the contamination from the dominant background of photons
originating from hadron decays.
As will be discussed later, imposing an isolation cut for the photons also reduces the fragmentation contribution
so that the prompt photon cross section will be more sensitive to the direct component.

The production of photons in heavy-ion collisions~\cite{Adare:2008ab,Afanasiev:2012dg,Adare:2012vn,Muller:2015taa,Morrison:2015waa,Yang:2014mla,Chatrchyan:2012vq,Wilde:2012wc,Adam:2015lda,Aad:2015lcb,Novitzky:2016lxl} looks a promising future tool for studying
the cold nuclear matter effects~\cite{Dai:2013xca,Helenius:2013bya}, since photons are not accompanied by any 
final state interaction and hence leave the system with their energy and momenta unaltered. 
It has also been recognised as a powerful tool to study the fundamental properties of
quark gluon plasma (QGP) created in these collisions~\cite{Shuryak:1980tp,d'Enterria:2006su,JalilianMarian:2001ey,Kopeliovich:2007sd,
KumarA:2013lwa,Mandala:2013xma,Singh:2015hqa}.
Furthermore, since the nuclear parton distribution functions (nPDFs)~\cite{Hirai:2007sx,Eskola:2009uj,deFlorian:2011fp,Khanpour:2016pph,Kovarik:2015cma,Hirai:2016ykc} (especially
of the gluon) cannot be well determined using the available nuclear deep inelastic scattering (DIS) and 
Drell-Yan experimental data compared with the PDFs of the free nucleon, 
the measurements of prompt photon production in heavy-ion collisions
can be used to constrain the gluon distributions within nuclei~\cite{Arleo:2007js,BrennerMariotto:2008st,Arleo:2011gc,Helenius:2014qla}.
One of the important questions in the theoretical calculation of the particle production cross sections 
in nuclear collisions is that whether the factorization theorem~\cite{Collins:1989gx,Brock:1993sz,Celmaster:1981dv,Nayak:2009ix} 
of collinear singularities is valid or not in this case (note that it is established in the case of hadronic collisions). 
So, the production of photons in nuclear collisions can also be recognised as a useful tool to answer
this question.

Although in Ref.~\cite{d'Enterria:2012yj}, the authors found a small effect on the gluon density
due to the inclusion of large number of isolated prompt photon production data until 2012 related to the various 
experiments at different center-of-mass energies in a global analysis of PDFs, it is expected that the
recent ATLAS data~\cite{Aad:2016xcr} measured at center-of-mass energy $ \sqrt s=8 $ TeV can be used to improve PDF fits especially at larger Bjorken scaling variable $ x $ where the PDF uncertainties are relatively large~\cite{Schwartz:2016olw}. 
Such expectation can be accounted for near future ATLAS measurements at 13 TeV~\cite{ATLAS13TeV}.
In this work, we are going to make predictions for the isolated prompt photon production in $ pp $ collisions
at $ \sqrt s=13 $ TeV using various modern PDF sets~\cite{Dulat:2015mca,Harland-Lang:2014zoa,Ball:2014uwa}.

The paper is organised as follows. In Sec.~\ref{sec:two}, we first 
describe briefly the prompt photon physics and introduce various 
prescription of photon isolation. Then, using various modern PDF sets,
we present the theoretical predictions for the isolated prompt photon production
at 13 TeV to study the impact of input PDFs on the obtained results.
The differential cross sections are presented both as
a function of $ E_\textrm{T}^\gamma $ and photon pseudorapidity $ \eta^\gamma $.
In Sec.~\ref{sec:three}, we study in detail the theoretical uncertainty
in the cross sections due to the variation of the renormalization, factorization and fragmentation scales
and determine its order of magnitude. In Sec.~\ref{sec:four}, we introduce and calculate the ratios of photon momenta 
for different rapidity regions and study the impact of various input PDFs on such quantity. Finally, our results and conclusions
are summarized in Sec.~\ref{sec:five}.

%
\section{Predictions for the isolated prompt photon production at 13 TeV}\label{sec:two}
Theoretical and computational aspects of the inclusive isolated prompt photon production
such as involved leading order (LO) and next-to-leading order (NLO) subprocesses, direct and fragmentation 
component of the cross section and photon isolation requirement have been discussed in many papers 
(for instance see Ref.~\cite{Catani:2002ny,Aurenche:2006vj}). Generally, the prompt photon cross section can be
calculated by convolving nonperturbative PDFs and FFs with a perturbative partonic cross section
by virtue of the factorization theorem. Actually, as mentioned in the Introduction, there are two
components contributing to the prompt photon cross section: direct and fragmentation parts. 
In view of the theoretical calculations, they can be computed in separate, though
they cannot be measured separately in the experiments. Accordingly,
the prompt photon cross section in hadronic collisions can be written as follows:
\begin{equation}
d\sigma^{\gamma +X}= d\sigma_{\textrm{dir}}^{\gamma +X}+d\sigma_{\textrm{fragm}}^{\gamma +X},
\label{eq1}
\end{equation}
where the first and second terms represent the direct and fragmentation contributions,
respectively, and $ X $ indicates the inclusive nature of the cross section as usual.

There are three scales that should be set in the calculation of the cross section Eq.~\eqref{eq1}.
For the direct part, the renormalization scale $ \mu $ appears in perturbative partonic cross section 
while the (initial state) factorization scale $ M $ appears both in partonic cross section and PDFs.
For the fragmentation part, in addition to $ \mu $ and $ M $, the partonic cross section
includes also the fragmentation scale $ M_F $ (final state factorization scale for the fragmentation process).
In this case, $ M_F $ also appears in the parton-to-photon fragmentation functions.
Note that whether for direct or fragmentation components, the renormalization scale $ \mu $
appears in the strong coupling constant $ \alpha_s $. In theoretical calculations
of the prompt photon production, some uncertainties come from 
scale variations. We study in detail these uncertainties
for the isolated prompt photon production at 13 TeV in the next section.

At LO, there are two Born-level subprocesses contributing to
the prompt photon production cross section: the quark-gluon Compton scattering $ q(\bar q)g \rightarrow \gamma q(\bar q) $
or quark-antiquark annihilation $ q\bar q \rightarrow \gamma g $. Although at NLO
there are more contributing subprocesses $ q(\bar q)g \rightarrow \gamma gq(\bar q) $ and 
$ q\bar q \rightarrow \gamma gg $ and the others from the virtual corrections to the Born-level processes,
the point-like coupling of the photon to quarks makes the 
calculations easier~\cite{Aurenche:1983ws,Aurenche:1987fs,Fontannaz:2001ek}
(note that the first calculation of direct photon production at next-to-next-to leading order
(NNLO) accuracy in QCD has also been presented recently~\cite{Campbell:2016lzl}).
It is established that the $ q\bar q $ annihilation channel is suppressed compared to the other subprocesses
at $ pp $ colliders such as LHC and RHIC whereas at the Tevatron that is a 
$ p\bar p $ collider, this channel is relevant~\cite{Ichou:2010wc}.

For measuring the prompt photon production at hadron colliders inclusively,
the background of secondary photons coming from the decays of hadrons produced in the collision
should be well rejected. We can do it by imposing appropriate isolation cuts.
As mentioned, the photon isolation also significantly reduces the fragmentation components
of the prompt photon cross section. Actually, the reason is that 
the fragmentation photons are emitted collinearly to the parent parton,
and on the other hand, the isolation cut discards the prompt photon events that have too much hadronic activity.
Here we introduce two prescription of photon isolation 
used so far in photon production studies. The most used is the cone criterion~\cite{Catani:2002ny}
that is defined as follows. A photon is isolated if, inside a cone of radius $ R $ centered around the photon
direction in the rapidity $ y $ and azimuthal angle $ \phi $ plane, the amount
of hadronic transverse energy $ E_\textrm{T}^{\textrm{had}} $ is smaller than some value
$ E_\textrm{T}^{\textrm{~max}} $:
\begin{equation}
E_\textrm{T}^{\textrm{had}}\leq E_\textrm{T}^{\textrm{~max}}, \qquad (y-y_\gamma)^2+(\phi-\phi_\gamma)^2\leq R^2.
\label{eq2}
\end{equation}
Although both the CMS and ATLAS Collaborations take $ R=0.4 $, the value of $ E_\textrm{T}^{\textrm{~max}} $
is different in their various measurements. For example, it is a finite value 5 GeV in the CMS measurement~\cite{Chatrchyan:2011ue} 
or 7 GeV in the ATLAS measurement~\cite{Aad:2013zba} both at $ \sqrt s=7 $ TeV whereas it has been considered
as a function of photon transverse energy $ E_\textrm{T}^\gamma $ as $ E_\textrm{T}^{\textrm{~max}} = 4.8~\textrm{GeV} + 0.0042~E_\textrm{T}^\gamma $
in the recent ATLAS measurement at $ \sqrt s=8 $ TeV~\cite{Aad:2016xcr}.
In another prescription of photon isolation proposed by S. Frixione~\cite{Frixione:1998hn}, 
the fragmentation components is suppressed while the cross section is kept infrared safe at any order in perturbative QCD.
In this case, the amount of $ E_\textrm{T}^{\textrm{had}} $ is required to satisfy the condition $ E_\textrm{T}^{\textrm{had}}\leq f(r) $,
for all radii $ r $ inside the cone described in Eq.~\eqref{eq2}. The energy profile function $ f(r) $ 
can be considered as 
\begin{equation}
f(r)=\epsilon_s E_\textrm{T}^\gamma \left( \frac{1-\cos(r)}{1-\cos(R)}\right)^n,
\label{eq3}
\end{equation}
where $ \epsilon_s $ and $ n $ are positive numbers of order one. Note that $ f(r) $ is an increasing function of $ r $
and falls to zero as $ r\rightarrow 0 $, since $ n $ is positive.

There are some computer codes can be used to calculate the prompt photon production
cross section at NLO such as J\textsc{et}P\textsc{hox}~\cite{Catani:2002ny,Aurenche:2006vj,Belghobsi:2009hx} 
and P\textsc{e}T\textsc{e}R~\cite{Becher:2013vva}. J\textsc{et}P\textsc{hox}
is a Monte Carlo programme written as a partonic event generator for the 
prediction of processes with photons in the final state. It can calculate
the direct and fragmentation contributions of the cross section, separately. The calculation can be configured
to specify several parameters like kinematic range, PDFs and FFs
and also to use an isolation cut with a finite value or $ E_\textrm{T}^\gamma $ dependent 
linear function for $ E_\textrm{T}^{\textrm{~max}} $ in Eq.~\eqref{eq2}.

Now we are in position to predict the isolated prompt photon photon in
$ pp $ collisions at center-of-mass energy of 13 TeV using various modern 
PDF sets (CT14~\cite{Dulat:2015mca}, MMHT14~\cite{Harland-Lang:2014zoa}, NNPDF3.0~\cite{Ball:2014uwa},
HERAPDF2.0~\cite{Abramowicz:2015mha} and JR14~\cite{Jimenez-Delgado:2014twa}).
In this way, we can also investigate the effect of the PDF choice on the predictions.
Note that for each group, its NLO PDF sets with $ \alpha_s(M_Z)=0.118 $ is taken
through the LHAPDF package~\cite{Buckley:2014ana}. 
It should be also noted that we use the kinematic settings 
introduced in Ref.~\cite{ATLAS13TeV}. All calculations in this work are performed 
using the J\textsc{et}P\textsc{hox} with including all
diagrams up to the LO and NLO order of QED and QCD coupling, respectively, defined in the
$ \overline{\textrm{MS}} $ renormalization scheme (It is worth pointing out in this context that
since the NNLO calculations~\cite{Campbell:2016lzl} have not yet been incorporated 
into any readily available codes like J\textsc{et}P\textsc{hox}, the NLO results are still interesting). 
The fine-structure constant ($ \alpha _{\textrm{EM}}$) 
is set to the J\textsc{et}P\textsc{hox} default of 1/137. Moreover, for
calculating the fragmentation component of the cross sections, we use in all predictions
the NLO Bourhis-Fontannaz-Guillet FFs of photons~\cite{Bourhis:1997yu}. 
The isolation transverse energy is taken to be  $ E_\textrm{T}^\gamma $ dependent 
as $ E_\textrm{T}^{\textrm{~max}} = 4.8~\textrm{GeV} + 0.0042~E_\textrm{T}^\gamma $~\cite{ATLAS13TeV}.
In all calculations that are
performed in this section, the renormalization ($ \mu $), factorization ($ M $) and fragmentation ($ M_F $ ) 
scales are setted to the photon transverse energy ($ \mu=M=M_F=E_\textrm{T}^\gamma $) and
the scale uncertainty is studied separately in the next section.

\begin{figure}[t!]
\centering
\includegraphics[width=8.6cm]{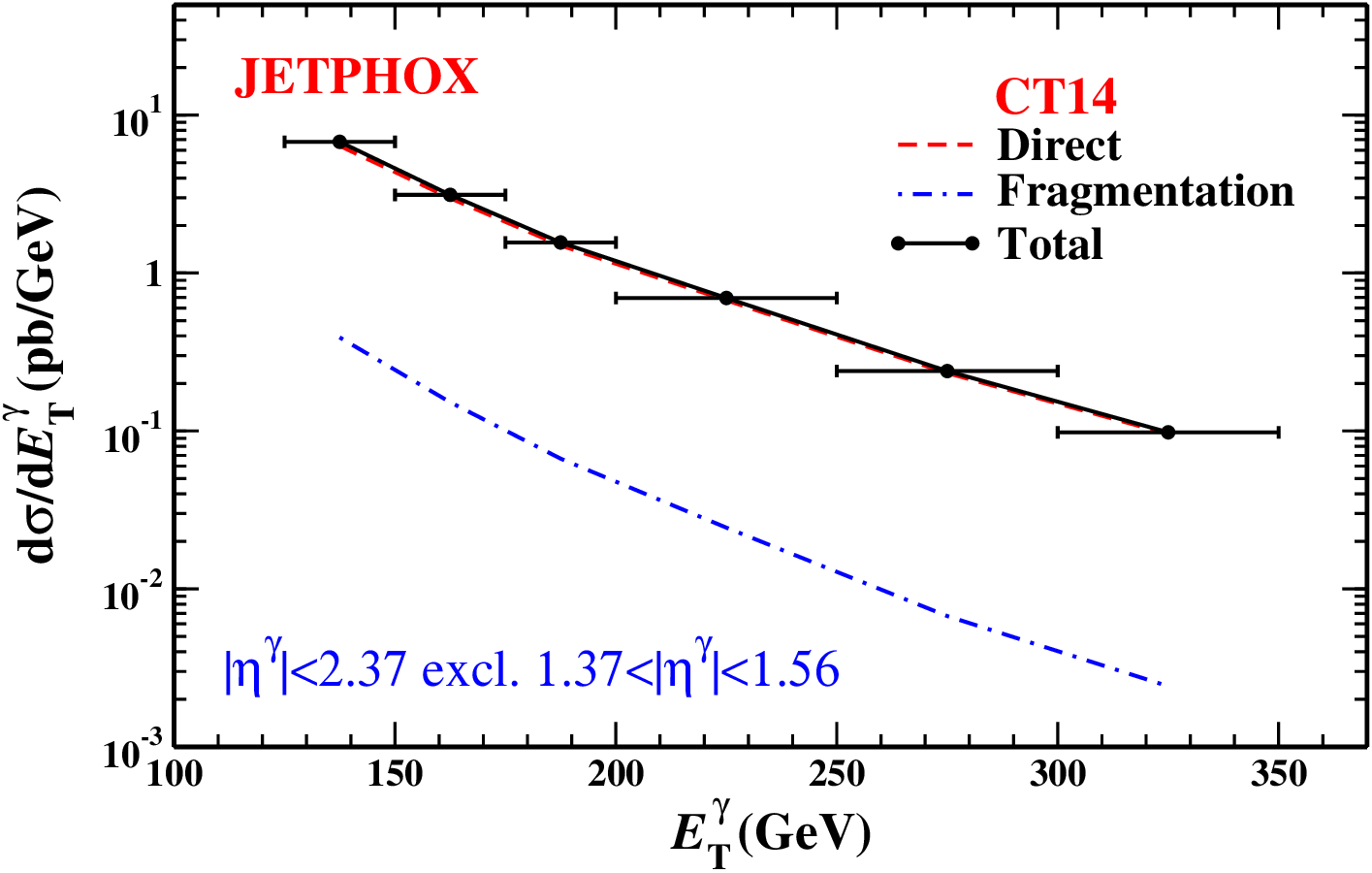}
\caption{The NLO differential cross section of the isolated prompt photon
production in $ pp $ collisions at $ \sqrt{s}=13 $ TeV as a function of $ E_\textrm{T}^\gamma $ 
in the kinematic range $ 125<E_\textrm{T}^\gamma<350 $ GeV for $ |\eta^\gamma|<2.37 $ excluding the region
$ 1.37<|\eta^\gamma|<1.56 $ and using NLO CT14~\cite{Dulat:2015mca} PDFs.
The direct (red dashed curve) and fragmentation (blue dotted-dashed curve) contributions to the 
total cross section (black solid curve) have been shown, separately.}
\label{fig:fig1}
\end{figure}

As a first step, we calculate the NLO differential cross section of the isolated prompt photon
production in $ pp $ collisions at $ \sqrt{s}=13 $ TeV as a function of $ E_\textrm{T}^\gamma $ 
in the kinematic range $ 125<E_\textrm{T}^\gamma<350 $ GeV for $ |\eta^\gamma|<2.37 $ excluding the region
$ 1.37<|\eta^\gamma|<1.56 $. It should be noted here that photons are detected in ATLAS by a lead-liquid Argon sampling
electromagnetic calorimeter (ECAL) with an accordion geometry,
divided into three sections: A barrel section covering the pseudorapidity region
$ |\eta^\gamma|<1.475 $ and two endcap sections covering the pseudorapidity
regions $ 1.375<|\eta^\gamma|<3.2 $. Measurement of the isolated prompt photon
production with the ATLAS detector is usually performed for $ |\eta^\gamma|<2.37 $ excluding the region
$ 1.37<|\eta^\gamma|<1.56 $ to include the detector region equipped with tracking detectors, 
but ignoring the transition region between the barrel
and endcap calorimeters where the detector response is not optimal~\cite{Aad:2011tw,Aad:2013zba,Aad:2016xcr}.
Fig.~\ref{fig:fig1} shows the obtained results using CT14 PDFs~\cite{Dulat:2015mca}
for direct (red dashed curve) and 
fragmentation (blue dotted-dashed curve) contributions to the cross section
and also total cross section (black solid curve), separately. 
Note that the horizontal error bars show the edges of each bin in $ E_\textrm{T}^\gamma $ 
and the theoretical uncertainties in the results are discussed separately in the next section.
This figure indicates that the direct component dominates completely the cross section, in all ranges
of $ E_\textrm{T}^\gamma $ especially at larger values. To be more precise, the
contribution of the fragmentation component to the total cross section is of the order of 5$ \% $ at smallest
value of $ E_\textrm{T}^\gamma $ and even less than 3$ \% $ at larger ones.
This fact can be very important in view of the phenomenology, because
we can use the future ATLAS data at $ \sqrt{s}=13 $ in a new global analysis of PDFs
without considering the fragmentation component, since its calculation can be time consuming
and also adds FFs uncertainties in the analysis (Note that our present knowledge of
photon fragmentation functions is not satisfactory enough).

\begin{figure}[!]
\centering
\includegraphics[width=8.6cm]{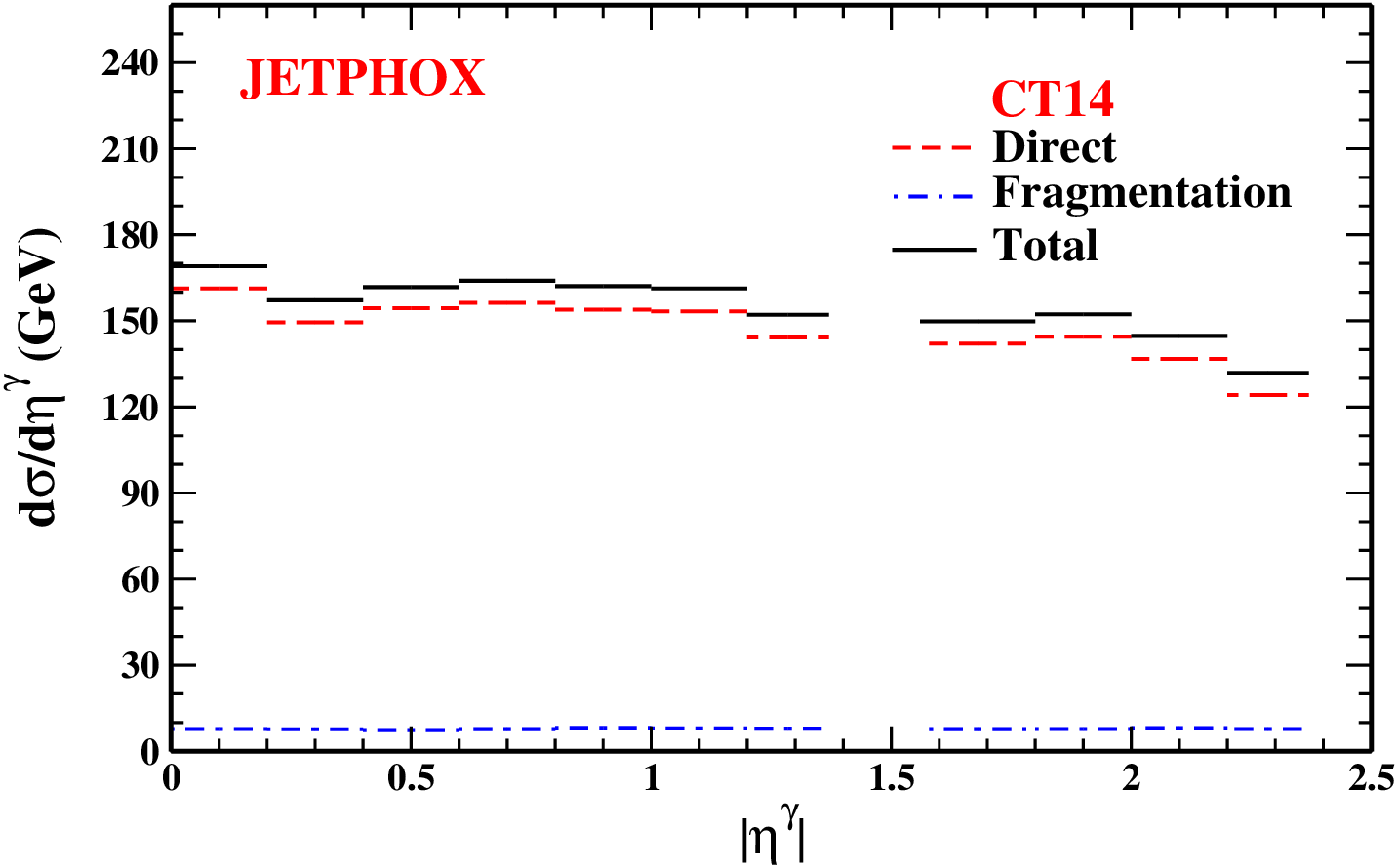}
\caption{Same as Fig. 1, but as a function of $ \eta^\gamma $.}
\label{fig:fig2}
\end{figure}

By virtue of the J\textsc{et}P\textsc{hox} facilities, we can also calculate 
the NLO differential cross section of the isolated prompt photon
production in $ pp $ collisions at $ \sqrt{s}=13 $ TeV as a function of photon pseudorapidity $ \eta^\gamma $.
The obtained results using CT14 PDFs for $ 125<E_\textrm{T}^\gamma <350 $ GeV and both $ |\eta^\gamma|<1.37 $ 
and $ 1.56<|\eta^\gamma|<2.37 $ regions have been shown in Fig.~\ref{fig:fig2} 
where we have again plotted both the direct (red dashed curve) and fragmentation (blue dotted-dashed curve) 
parts and also total cross section (black solid curve), for comparison. In this case, the
contribution of the fragmentation component to the cross section is either about 5$ \% $
at all values of $ \eta^\gamma $ and then completely negligible compared with the direct component.
\begin{figure}[t!]
\centering
\includegraphics[width=8.6cm]{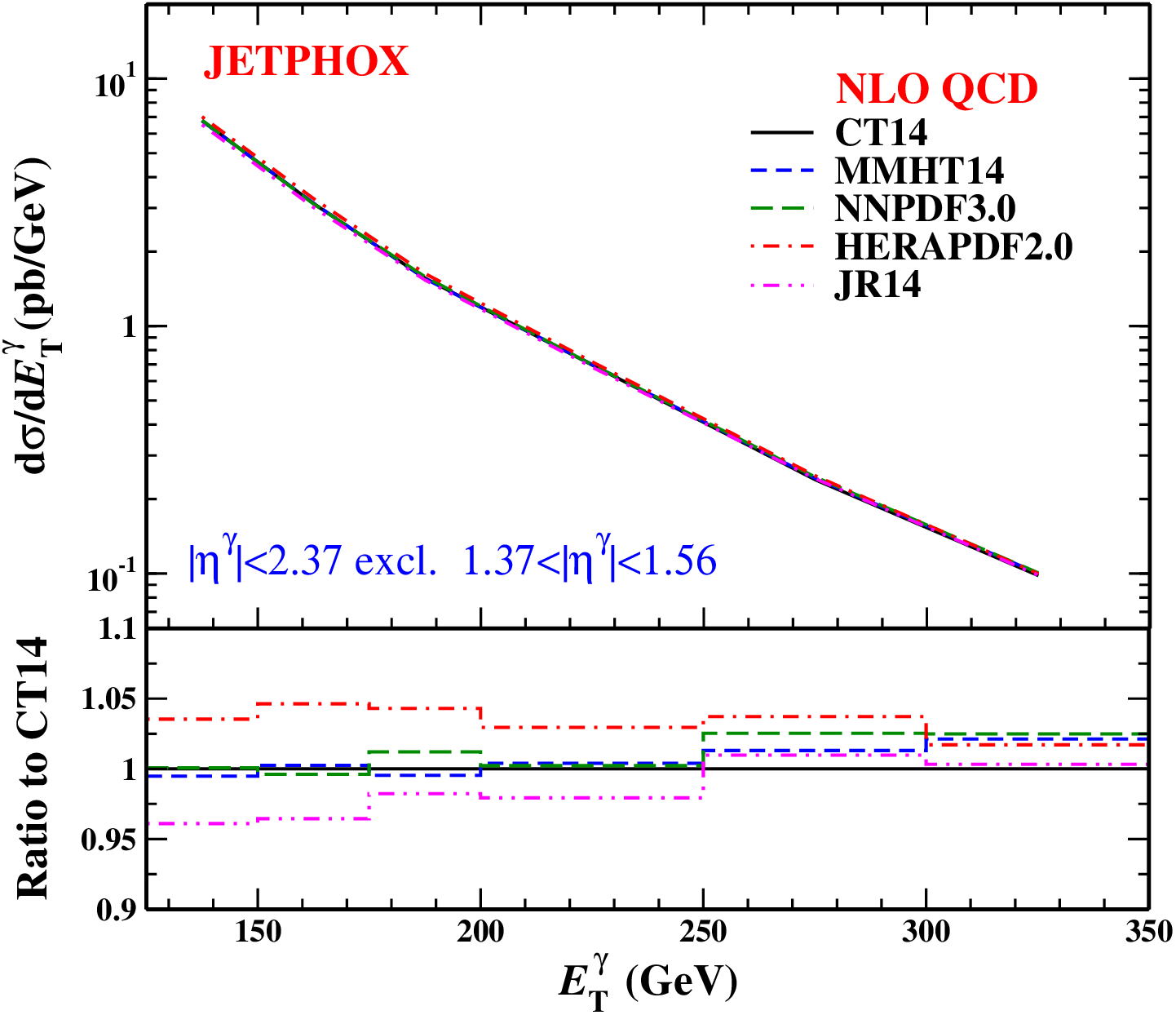}
\caption{A comparison of the NLO theoretical predictions for the total
differential cross section of the isolated prompt photon production as a function of $ E_\textrm{T}^\gamma $ 
using various NLO PDFs of CT14~\cite{Dulat:2015mca} (black solid curve),
MMHT14~\cite{Harland-Lang:2014zoa} (blue dashed curve),
NNPDF3.0~\cite{Ball:2014uwa} (green long-dashed curve), HERAPDF2.0~\cite{Abramowicz:2015mha} (red dotted-dashed curve) 
and JR14~\cite{Jimenez-Delgado:2014twa} (pink dotted-dotted-dashed curve) at $ \sqrt{s}=13 $ TeV
in the kinematic range $ 125<E_\textrm{T}^\gamma<350 $ GeV 
for $ |\eta^\gamma|<2.37 $ excluding the region $ 1.37<|\eta^\gamma|<1.56 $.
Ratio to the central value of CT14 has been shown in the bottom panel.}
\label{fig:fig3}
\end{figure}
\begin{figure}[!]
\centering
\includegraphics[width=8.6cm]{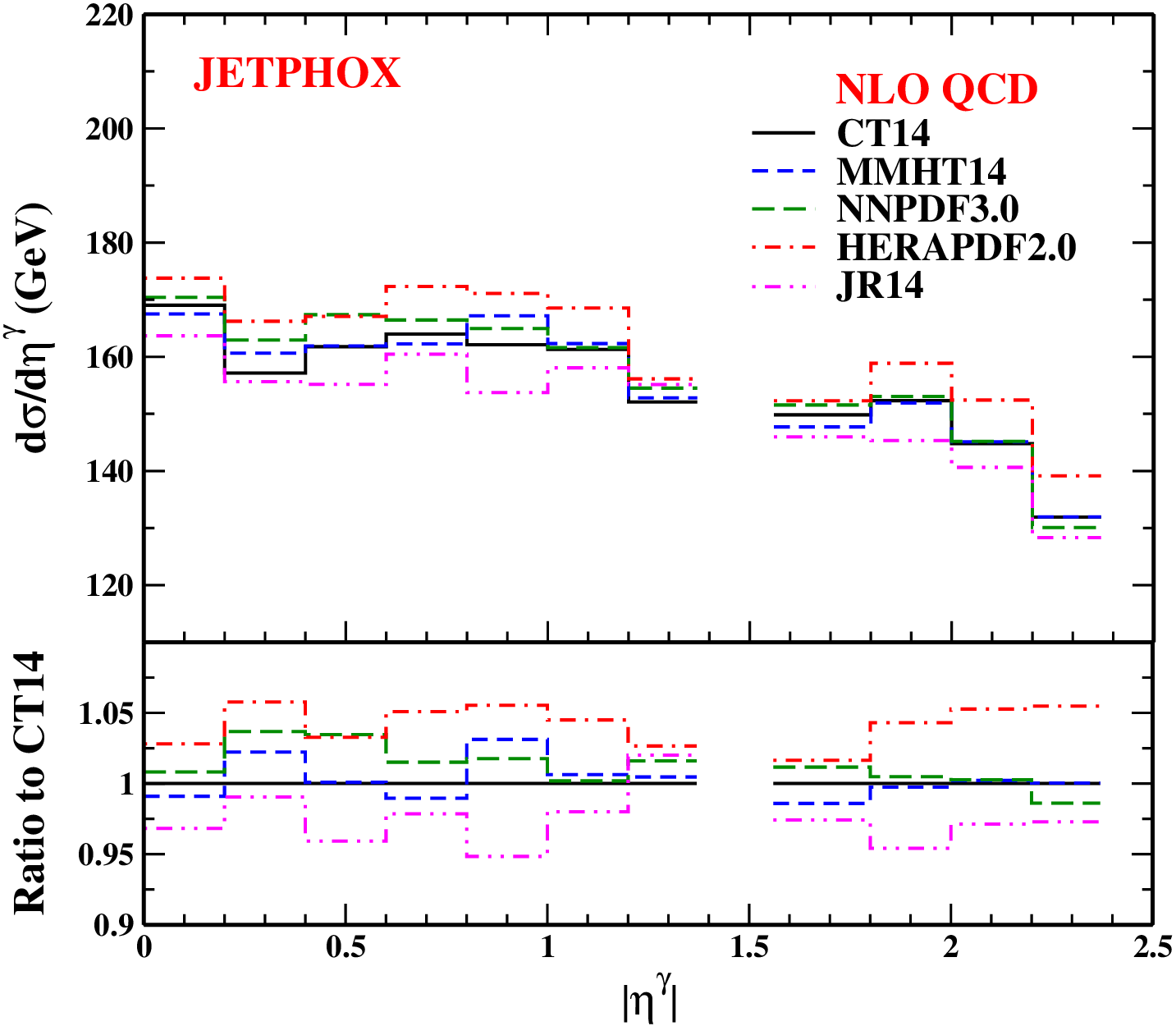}
\caption{Same as Fig. 3, but as a function of $ \eta^\gamma $.}
\label{fig:fig4}
\end{figure}
In order to study the impact of input PDFs on the final results and
estimate the order of magnitude of the difference between their predictions,
we can now recalculate the differential cross sections
presented in Figs.~\ref{fig:fig1} and~\ref{fig:fig2}, but this time using other PDF sets.
To this aim, we choose the NLO MMHT14~\cite{Harland-Lang:2014zoa}, NNPDF3.0~\cite{Ball:2014uwa},
HERAPDF2.0~\cite{Abramowicz:2015mha} and JR14~\cite{Jimenez-Delgado:2014twa}
PDF sets (It should be noted that we use the dynamical PDFs set of JR14). Figs.~\ref{fig:fig3} and~\ref{fig:fig4} show the comparison between their
predictions for the total differential cross section of the isolated prompt photon production
in $ pp $ collisions at $ \sqrt{s}=13 $ TeV as a function of $ E_\textrm{T}^\gamma $ and $ \eta^\gamma $
for the same kinematic settings as Figs.~\ref{fig:fig1} and~\ref{fig:fig2}, respectively.
The difference between the predictions in the various kinematic regions can be
investigated in more details from the bottom panel of each figure where 
the ratio of all predictions to the central value of CT14 have been shown.
As can be seen, for both cross sections, all predictions are in good agreement with each other so that,
for example, the CT14, MMHT14 and NNPDF3.0 are same to a large extent at smaller values of $ E_\textrm{T}^\gamma $ in Fig.~\ref{fig:fig3}.
However, the differences between the HERAPDF2.0 and JR14 predictions with CT14 are somewhat larger than
the others at low $ E_\textrm{T}^\gamma $. 
In overall, we can state that the difference between these PDF sets is up to 5$ \% $. 
This is due to the fact that the parton distributions from various PDF sets,
especially of the gluon in this case, become very similar at very high energies.
Note also that in view of the experimental uncertainties~\cite{ATLAS13TeV},
the total systematic uncertainty is smaller than 5\% at low values of $E_{\mathrm T}^\gamma $
and it increases as $E_{\mathrm T}^\gamma $ increases. Therefore, considering only the systematic uncertainty,
discrimination between the theoretical predictions at the level of 5\% is going to be possible just at low values of $E_{\mathrm T}^\gamma $.
However, although the systematic uncertainty dominates the total experimental uncertainty at 
low values of $E_{\mathrm T}^\gamma $, the statistical uncertainty should also be considered
as it increases towards high $E_{\mathrm T}^\gamma $.

%
\section{The study of scale uncertainty}\label{sec:three} 
In the previous section we calculated the cross
section of isolated prompt photon production in $ pp $ collisions using various PDF sets.
Now, it is important to calculate and study the theoretical uncertainties in the results.
Since the dominant theoretical uncertainty is that arising from the scale uncertainties, in
this section, we discuss only the scale uncertainties and ignore the study of PDFs 
uncertainties (note that the uncertainty arising from those in the PDFs amounts to 1-4\%).
As discussed in the previous section, the NLO calculation of the 
isolated prompt photon production involves all three 
renormalization ($ \mu $), factorization ($ M $) and fragmentation ($ M_F $ ) 
scales. If we could calculate the cross section to all orders in perturbation theory,
we could say that the cross section is scale independent and there is no  
theoretical uncertainty on the results due to the scales choice.
But, the scales choice become an important issue when we calculate the 
cross section to a fixed order in $ \alpha_s $. Since the mentioned scales are all unphysical, 
the more reliable predictions are those for which
the dependence of the cross section on the scales is minimised. It has been established that
no optimal scale choice is possible for the prediction of the inclusive 
photon cross section in the region of the phase space of interest~\cite{Blair}.
In this way, it was accepted that the predictions and their uncertainties should be made
by setting all scales to be equal and varying them of a factor 2 around the central value $ \mu=M=M_F=E_\textrm{T}^\gamma $.
However, if we want to be more correct in the calculation of the scale uncertainties, we
should follow a method consisting of the combination of both incoherent and coherent scales variation~\cite{Blair}.
To be more precise, in an incoherent variation one should vary the scales independently
by a factor of 2 around the central value so that one scale is varied keeping the other two equal to $ E_\textrm{T}^\gamma $.
In a coherent variation one should vary the scales simultaneously by a factor of 2 around the central value as before.
Then, the total scale uncertainty can be calculated by adding in quadrature all obtained uncertainties
considering following constraints:
\begin{itemize}
  \item $\mu = M = M_F \in[E_\textrm{T}^\gamma/2,  2E_\textrm{T}^\gamma]$;
  \item $\mu \in[E_\textrm{T}^\gamma/2, 2 E_\textrm{T}^\gamma], ~M = M_F = E_\textrm{T}^\gamma$;
  \item $M \in[E_\textrm{T}^\gamma/2, 2 E_\textrm{T}^\gamma], ~\mu = M_F = E_\textrm{T}^\gamma$;
  \item $M_F \in[E_\textrm{T}^\gamma/2, 2 E_\textrm{T}^\gamma], ~\mu = M = E_\textrm{T}^\gamma$.
\end{itemize} 
\begin{figure}[t!]
\centering
\includegraphics[width=8.6cm]{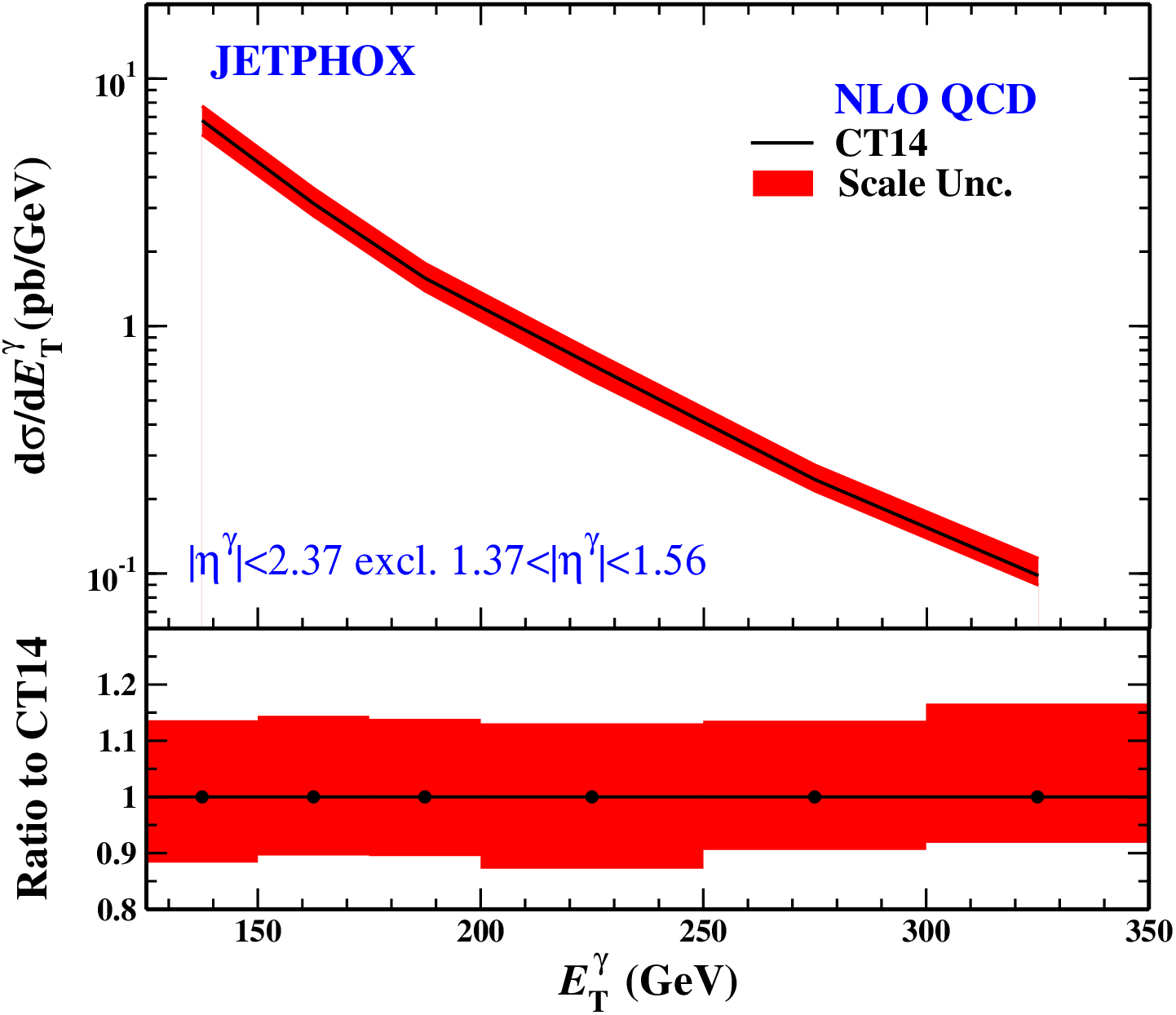}
\caption{The NLO theoretical predictions for the total differential cross section
of the isolated prompt photon as a function of $ E_\textrm{T}^\gamma $ using NLO CT14~\cite{Dulat:2015mca} PDFs with scale
uncertainty (red band) at $ \sqrt{s}=13 $ TeV
in the kinematic range $ 125<E_\textrm{T}^\gamma<350 $ GeV 
for $ |\eta^\gamma|<2.37 $ excluding the region $ 1.37<|\eta^\gamma|<1.56 $. 
Ratio to the central value of CT14 has been shown in the bottom panel.}
\label{fig:fig5}
\end{figure}
In order to study the scale uncertainty of the isolated prompt photon
production cross section in $ pp $ collisions at $ \sqrt{s}=13 $ TeV, we
again select the CT14~\cite{Dulat:2015mca} PDFs and perform the calculations both as a function of
$ E_\textrm{T}^\gamma $ and $ \eta^\gamma $ for the ATLAS kinematic~\cite{ATLAS13TeV}.
Figs.~\ref{fig:fig5} and~\ref{fig:fig6} show the obtained results where the predictions and
scale uncertainties have been shown as black solid curves and red bands, respectively.
The ratio to CT14 central prediction has been shown in the bottom panel of each figure.
As one can see, the scale uncertainty can reach 20$ \% $ in some regions.
The large scale variations indicate that the NNLO calculations are needed 
to make more realistic theoretical predictions. Such calculations~\cite{Campbell:2016lzl} are now
becoming available and will be the subject of further work.
\begin{figure}[!]
\centering
\includegraphics[width=8.6cm]{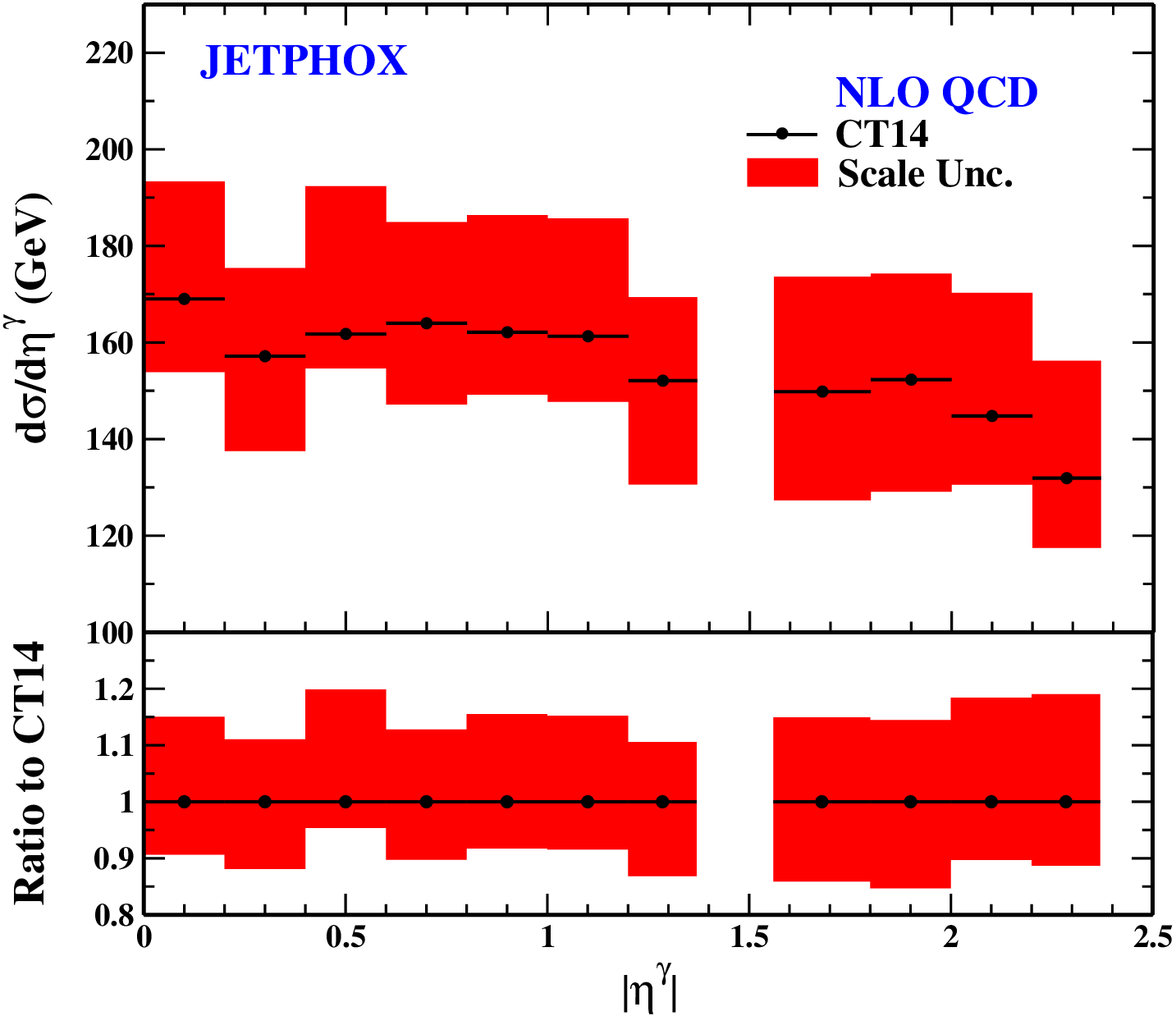}
\caption{Same as Fig. 5, but as a function of $ \eta^\gamma $.}
\label{fig:fig6}
\end{figure}
%

%
\section{the ratios of photon momenta for different rapidity regions}\label{sec:four}
\begin{figure}[!]
\centering
\includegraphics[width=8.6cm]{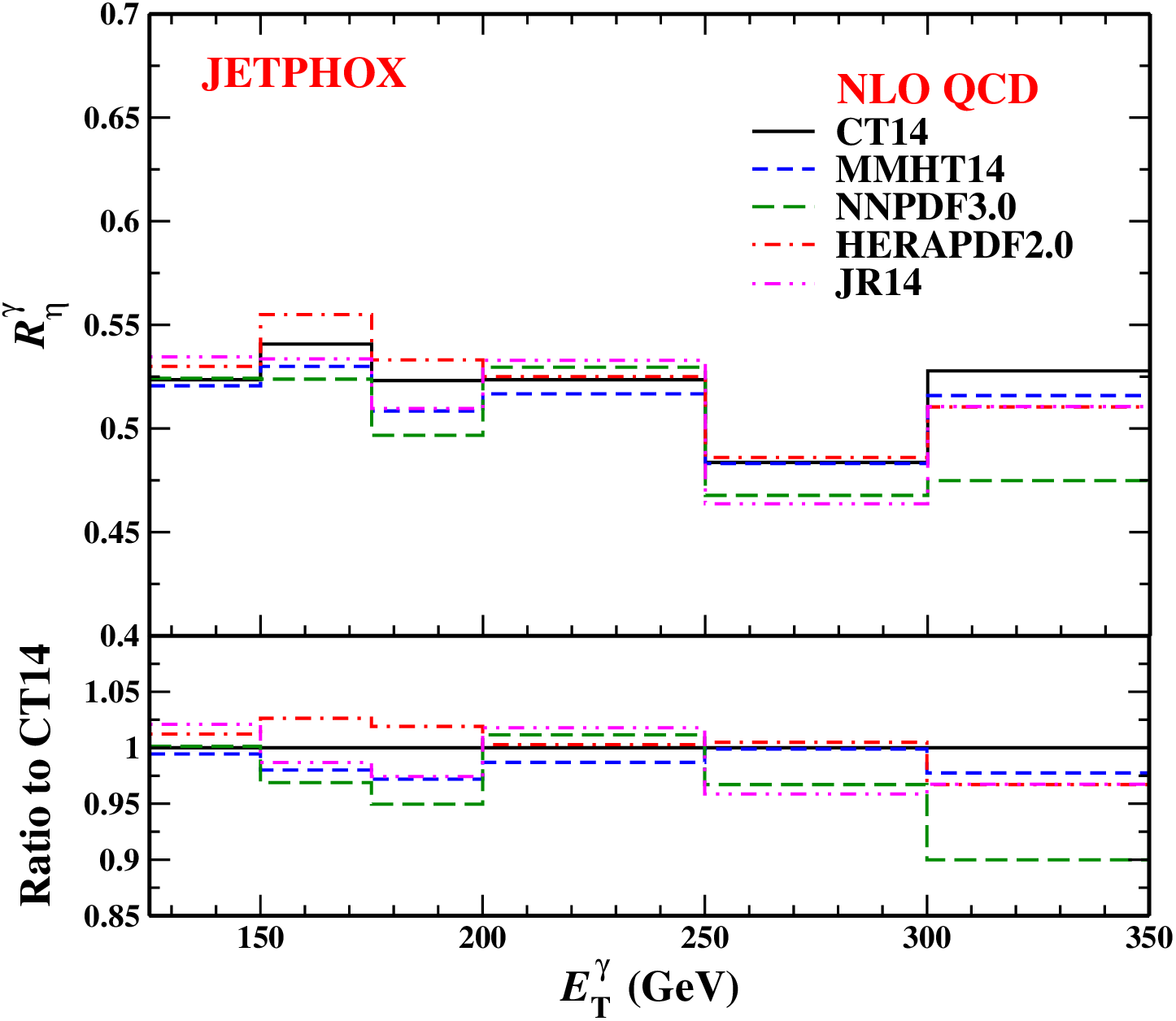}
\caption{A comparison of the ratio of the NLO theoretical predictions for the differential cross section
of the isolated prompt photon for the rapidity region $ 1.56<|\eta^\gamma|<2.37 $ to the same ones but
for the rapidity region $ |\eta^\gamma|<1.37 $ as a function of $ E_\textrm{T}^\gamma $
using various NLO PDFs of CT14~\cite{Dulat:2015mca} (black solid curve),
MMHT14~\cite{Harland-Lang:2014zoa} (blue dashed curve),
NNPDF3.0~\cite{Ball:2014uwa} (green long-dashed curve), HERAPDF2.0~\cite{Abramowicz:2015mha} (red dotted-dashed curve) 
and JR14~\cite{Jimenez-Delgado:2014twa} (pink dotted-dotted-dashed curve) at $ \sqrt{s}=13 $ TeV
in the kinematic range $ 125<E_\textrm{T}^\gamma<350 $ GeV 
for $ |\eta^\gamma|<2.37 $ excluding the region $ 1.37<|\eta^\gamma|<1.56 $.
Ratio to the central value of CT14 has been shown in the bottom panel.}
\label{fig:fig7}
\end{figure}

As we saw in the previous section, if one consider the combination of both incoherent and coherent scale variations, 
the resulting scale uncertainty is considerably large. Generally, the decrease of the total uncertainty
origination from various sources is a very important issue both in the experimental measurements and
theoretical calculations. In most cases, the expression of results as ratios can be very useful to this aim.
For example, in nuclear collisions, it is well established now that the measurement of nuclear modification and forward-to-backward
ratios is more suitable than single differential cross section~\cite{Arleo:2011gc,Helenius:2014qla}.
In this section, we calculate and study the ratios of photon momenta 
for different rapidity regions using various input PDFs. Such ratios have the advantage of cancelling
some theoretical and experimental uncertainties. Consider the relation
\begin{equation}
R_{\eta}^\gamma \equiv \frac{d\sigma/dE_\textrm{T} \mid_{\eta\in[\eta_1,\eta_2]}}{d\sigma/dE_\textrm{T} \mid_{\eta\in[\eta_3,\eta_4]}},
\label{eq4}
\end{equation}
in which $ [\eta_1,\eta_2] $ and $ [\eta_3,\eta_4] $ represent different rapidity regions.
Note that since the differential cross section is sensitive the different values of $ x $
in different rapidity regions, then $ R_{\eta}^\gamma $ can probe the input PDFs
in a more curious way. Now, we calculate the ratios of the NLO theoretical predictions for the differential cross section
of the isolated prompt photon for the rapidity region $ 1.56<|\eta^\gamma|<2.37 $ to the same ones but
for the rapidity region $ |\eta^\gamma|<1.37 $. The calculations are performed again
using NLO PDFs of CT14~\cite{Dulat:2015mca},
MMHT14~\cite{Harland-Lang:2014zoa}, NNPDF3.0~\cite{Ball:2014uwa}, HERAPDF2.0~\cite{Abramowicz:2015mha}
and JR14~\cite{Jimenez-Delgado:2014twa} at $ \sqrt{s}=13 $ TeV. Fig.~\ref{fig:fig7}
shows the obtained results as a function of $ E_\textrm{T}^\gamma $. The
ratio to the central value of CT14 has been shown in the bottom panel. Compared
with Fig.~\ref{fig:fig3} (see the bottom panel of two figures), the difference between the HERAPDF2.0 and JR14 predictions
with the CT14 decreases at low values of $ E_\textrm{T}^\gamma $ in this case. However,
the NNPDF3.0 prediction is taken away from CT14 towards larger values of $ E_\textrm{T}^\gamma $
so that the difference between them is reached even to 10\%.
%

%
\section{SUMMARY and conclusions}\label{sec:five}
The study of the energetic photons produced in the collision of two hadrons provides 
a probe of perturbative QCD and can also give us some valuable information
about the parton densities in the nucleon and nuclei especially of the gluon. 
Photon production in heavy-ion collisions is also a powerful tool to study 
the cold nuclear matter effects and the fundamental properties of QGP. 
It is indicated that the recent ATLAS data~\cite{Aad:2016xcr} measured at center-of-mass 
energy $ \sqrt s=8 $ TeV can be used to improve PDF fits especially at larger Bjorken scaling variable $ x $~\cite{Schwartz:2016olw}.
So, the near future ATLAS measurement at 13 TeV~\cite{ATLAS13TeV} has more important
role in this respect. In the present paper, we presented the theoretical predictions 
for the isolated prompt photon production in $ pp $ collisions at $ \sqrt{s}=13 $ TeV 
both as a function of photon transverse energy $ E_\textrm{T}^\gamma $ and 
pseudorapidity $ \eta^\gamma $. All calculations performed 
using the J\textsc{et}P\textsc{hox} with including all
diagrams up to the LO and NLO order of QED and QCD coupling, respectively, defined in the
$ \overline{\textrm{MS}} $ renormalization scheme.
The isolation transverse energy is taken to be  $ E_\textrm{T}^\gamma $ dependent 
as $ E_\textrm{T}^{\textrm{~max}} = 4.8~\textrm{GeV} + 0.0042~E_\textrm{T}^\gamma $~\cite{ATLAS13TeV}.
As a result, we found that the direct component 
dominates completely the cross section in both cases, so that the
contribution of the fragmentation component to the total cross section is not more than 5$ \% $
and is even reduced to 3$ \% $ at some regions. So, 
we can study the impact of future ATLAS data at $ \sqrt{s}=13 $ on PDFs in a new global analysis,
neglecting the fragmentation component since its calculation can be time consuming
and also adds FFs uncertainties in the analysis. Then we compared 
the predictions from various modern PDF sets
namely the CT14~\cite{Dulat:2015mca}, MMHT14~\cite{Harland-Lang:2014zoa}, 
NNPDF3.0~\cite{Ball:2014uwa}, HERAPDF2.0~\cite{Abramowicz:2015mha} and JR14~\cite{Jimenez-Delgado:2014twa}
to investigate the effect of the PDF choice on the cross sections.
We found that all predictions are good agreement with each other.
To be more precise, in overall, the greatest difference between them is about 5$ \% $.
This can be attributed to the similarity of the parton distributions,
especially of the gluon in this case, from various PDF sets at very high energies.
In particular, the CT14, MMHT14 and NNPDF3.0 predictions are same to a large extent at smaller values of $ E_\textrm{T}^\gamma $ while
the HERAPDF2.0 and JR14 predictions differ a little more with them.
We also studied in detail the theoretical uncertainty
in the cross sections due to the variation of the renormalization, factorization and fragmentation scales.
The method consists of the combination of both incoherent and coherent scales variation.
We found that the scale uncertainty can reach 20$ \% $ in some regions so
the NNLO calculations are needed to make more realistic theoretical predictions. Finally,
we calculated the ratios of photon momenta for different rapidity regions and studied 
the impact of various input PDFs on such quantity. Is has the advantage of cancelling
some theoretical and experimental uncertainties and can probe the input PDFs
in a more curious way because the differential cross section is sensitive the different values of $ x $
in different rapidity regions.

%

\end{document}